\documentclass[12pt]{article}
\usepackage{graphics,epsfig}

\begin{document}

\title{Killer Geometries in Competing Species Dynamics}
\author{
{\bf Serge Galam\footnote{galam@ccr.jussieu.fr}}\\
 Laboratoire des Milieux D\'esordonn\'es et
H\'et\'{e}rog\`enes\thanks{Laboratoire associ\'e au CNRS (UMR
n$^{\circ}$
800) et \`a l'Universit\'e P. et M. Curie - Paris 6},\\ Tour 13 - Case
86,
4 place Jussieu, \\ 75252 Paris Cedex 05, France\\
\\     {\bf  Bastien Chopard\footnote{Bastien.Chopard@cui.unige.ch}}\\
D\'epartement d'Informatique, University of Geneva,\\
 24 rue G\'en\'eral-Dufour, 1211 Gen\`eve 4, Switzerland,\\
\\     {\bf Michel Droz\footnote{Michel.Droz@physics.unige.ch}}\\
D\'epartement de Physique Th\'eorique, University of Geneva,\\
 24 quai Ernest-Ansermet, 1211 Gen\`eve 4, Switzerland,}

\date{$\ $}
\maketitle

\begin{abstract}
We discuss a cellular automata model to study the competition
between an emergent better fitted species against an existing majority species..
The model implement local fights among small group of individual and a synchronous
random walk on a 2D lattice. The faith of the system, i.e. the
spreading or disappearence of the species
 is determined by their initial density and fight frequency. The
initial density of the emergent species has to be higher than a
critical threshold for total spreading but this value depends in a
non-trivial way of the fight frequency.  Below the threshold any
better adapted species disappears showing that a qualitative advantage
is not enough for a minority to win. No strategy is involved but
spatial organization turns out to be crucial. For instance at minority
densities of zero measure some very rare local geometries which occur
by chance are found to be killer geometries. Once set they lead with
high probability to the total destruction of the preexisting majority
species. The occurrence rate of these killer geometries is function of
the system size. This model may apply to a large spectrum of competing
groups like smoker-non smoker, opinion forming, diffusion of
innovation setting of industrial standards, species evolution,
epidemic spreading and cancer growth.

\end{abstract}

PACS numbers: 87.23.Kg, 05.50+q, 64.60-i

\section{Introduction}

The social behavior of a group of persons is certainly related in part to the
fact that each individual has its own autonomy and perception of the
environment.  However the global behavior may also reflect some
``mechanical'' response of each individual to the specific situation
it is confronted with. Such a collective behavior may possibly be
captured at by some cellular automaton model (see~\cite{BC-livre} for a general
introduction to cellular automata) provided the rules, to which each
individual obeys, are suitably chosen.

Here we address the generic problem of the competing fight between two
different groups over a fixed area.  We present a ``voter model''
which describes the dynamical behavior of a population with bimodal
conflicting interests and study the conditions of extinction of one of
the initial groups~\cite{galam:98, chopard:00}. Note that other interesting
applications can be addressed with this model, such as the problem of cancer tumor
growth~\cite{galam:01}.

\section{The Model}

Our model can be illustrated by the smoker - non smoker confrontation.
In a small group of persons a majority of smokers will usually
convince the few others to allow them to smoke making smoking the non
smokers at least passively and vice versa. But each time an equal
number of smokers and non-smokers meet an incertainty occurs. In that
case it may be assumed that a social trend will decide between the two
attitudes. In the US, smoking is viewed as a overall disadvantage
whereas in France it is rather well accepted. In other words, there is
a bias that will select the winner attitude in an even situation. In
our example, whether one studies the French or US case, the bias is in
favor of the smokers or the non-smokers, respectively. In our model such a smoking
non-smoking choices is considered for each new social encounter.

The same mechanism can be associated with the problem of two competing
standards. The choice of one or the other standard
is often driven by the opinion of the majority of people one meets. But
when the two competing systems are equally represented, the intrinsic
quality of the product becomes decisive. In that case, price and technological
advantage play the role of a bias.

Here we consider the simpler case of four-person encounters in a
spatially extended system in which the actors (species $A$ or $B$)
move randomly.  Initially, the $B$ species is present with density
$b_0$ and the $A$ species with density $1-b_0$. The $B$ individuals
are supposed to have a qualitative advantage over the $A$s but are
less numerous. The question we want to address is what is the minimal
density $b_0$ which make the $B$s win over $A$ (i.e. invade the entire
system at the expense of $A$ individuals).  Thus this model represents a
process of spatial contamination of opinion. Continuous approach have been also 
considered ~\cite{7}

The CA rule we propose here~\cite{galam:98} to describe this proceess is derived
from a model by Galam~\cite{galam}, in which the four individuals involved in a
tournament are randomy chosen among the current population, whose
composition in $A$ or $B$ type of person evolves after each
confrontation.  The density threshold for an invading emergence of $B$
is $b_c=0.23$ if the $B$ group has a qualitative bias over
$A$. However, with a spatial distribution of the species, even if
$b_0<b_c$, $B$ can still win over $A$ provided that it strives for
confrontation.  However, when the qualitative advantage is  not 
enough to win, a geographic as well as a definite degree of
aggressiveness is  instrumental to overcome the less fitted majority.

The model we use to describe the two populations $A$ and $B$ influencing
each other or competing for some unique resources, is based on the
diffusion automaton proposed in~\cite{BC-livre,galam:98}.
Particles have two possible
internal states ($\pm1$), coding for the $A$ or $B$ species,
respectively. Individuals move on a two-dimensional square lattice. At each site,
there are always four individuals (any combination of $A$'s and $B$'s is
possible). These four individuals all travels in a different lattice
direction (north, east, south and west). The situation is illustrated in
Fig. 1

\begin{figure}
\centerline{\psfig{figure=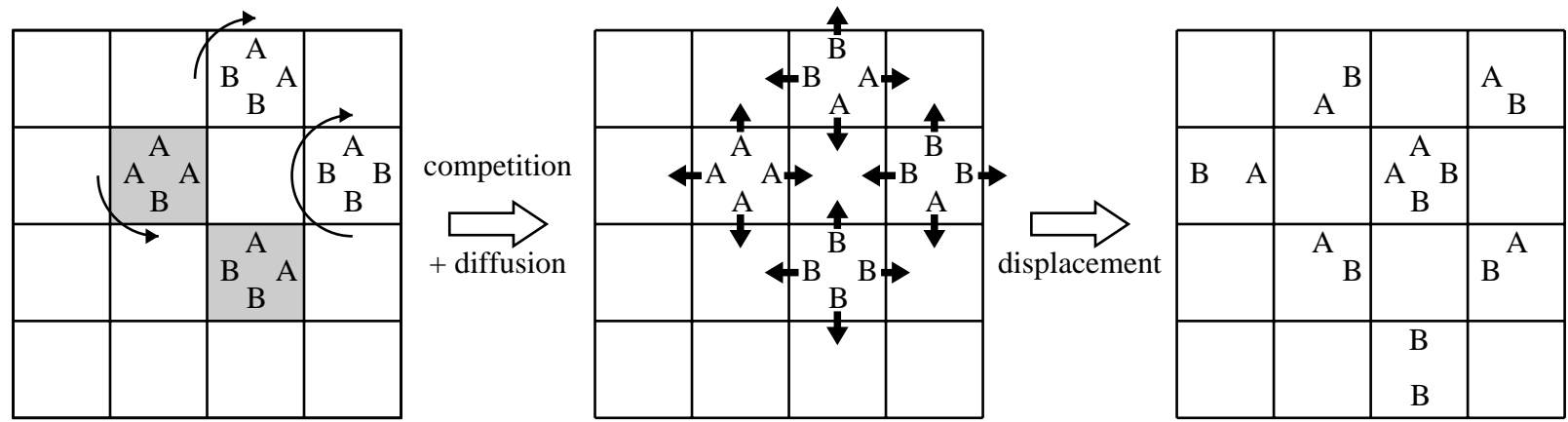}}
\caption{Sketch of the model rule. The symbols $A$ and $B$ denote the
two types 
of individuals. A confrontation take place in all gray cells and
results in a local victory of one species. Then, in all cells a random
re-direction of the individuals is performed (with a rotation of the
configuration by 0, 90, -90 or 180 degrees), followed by a jump to the
nearest neighbor cell.}
\end{figure}

The interaction takes place in the form of ``fights'' between the four
individuals meeting on the same site. At each fight, the group nature
($A$
or $B$) is updated according to the majority rule when possible,
otherwise
with a bias in favor of the best fitted group. The rules are,
\begin{itemize}
\item The local majority species (if any) wins:
\[ nA+mB\rightarrow \left\{ \begin{array}{ll}
                        (n+m)A & \mbox{if $n>m$} \\
                        (n+m)B & \mbox{if $n<m$} \\
                             \end{array}
\right.
\]
where $n$ and $m$ are integers satisfying the constraint $n+m=4$. 

\item When there is an equal number of $A$ and $B$ on a site, $B$ wins
the confrontation with probability $1/2+\beta/2$. The quantity
$\beta\in[0,1]$ is the bias accounting for some advantage (or extra
fitness) of species $B$.
\end{itemize}

Above rule is applied with probability $k$. Thus, with probability
$1-k$ the group composition does not change because no fight occurs.
Between fights both population agents diffuse on the
lattice, by randomly choosing a new direction (see~\cite{BC-livre} for more detail).

\begin{figure}
\centerline{\psfig{figure=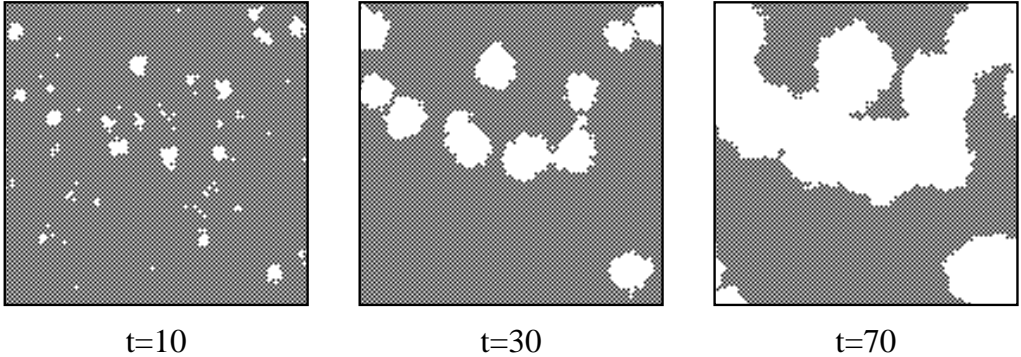,width=14cm}}
\caption{ Configurations of the CA model, at three different
times. 
The $A$ and $B$ species are represented by the gray and white regions,
respectively. The parameters of the simulation are $b_0=0.1$,
$k=0.5$ and $\beta=1$.}
\label{fig-smoker-snapshot}
\end{figure}

The behavior of this model is illustrated in
Fig.~\ref{fig-smoker-snapshot}.
The current configuration is shown at three different time steps. We can
observe the growth of dense clusters of $B$ invading the system.

It is clear  that the model richness comes from the even confrontations. If only odd
fights would happen, the initial majority population would always win after some
short time. The key parameters of this model are (i) $k$, the aggressiveness
(probability of confrontation), (ii) $\beta$, the $B$'s bias of winning
a
tie and (iii) $b_0$, the initial density of $B$.

The strategy according to which a minority of $B$'s (with yet a
technical,
genetic, persuasive advantage) can win against a large population of
$A$'s
is not obvious. Should they fight very often, try to spread or accept a
peace agreement? We study the parameter space by running the cellular
automaton.

In the limit of low aggressiveness ($k\to 0$), the particles move a long
time before fighting. Due to the diffusive motion, correlation between
successive fights are destroyed and, for $\beta=1$, $B$ wins provided that $b_0>0.23$.
 This is the mean-field level of our dynamical model which corresponds
to the theoretical calculations made in~\cite{galam}.

More generally, we observe that $B$ can win
even when $b_0<0.23$, provided it acts aggressively, i.e. by having a
large
enough $k$. Thus, there is a critical density $b_{death}(k)<0.23$ such
that, when $b_0>b_{death}(k)$, all $A$ are eliminated in the final
outcome.
Below $b_{death}$, $B$ looses unless some specific spatial
configurations
of $B$'s are present.

Therefore the growth of species
$B$ at the expense of $A$ is obtained by a spatial organization. Small
clusters that may accidentally form act as nucleus from which the $B$'s
can develop.  In other words, above the mean-field threshold $b_c=0.23$
there is no need to organize in order to win but, below this value only
condensed regions will be able to grow. When $k$ is too small, such an
organization is not possible (it is destroyed by diffusion) and the
strength advantage of $B$ does not lead to success.

Figure 3 summarizes, as a function of $b_0$ and $k$, the regions where
either $A$ or $B$ succeeds. It is found by inspection that the
separation curve satisfies the equation $(k+1)^7(b_0-0.077)=0.153$.

It is also interesting to study the time needed to annihilate completely
the looser. Here, time is measured as the number of fights per site
(i.e.
$kt$ where $t$ is the iteration time of the automaton).  We observe that
the dynamics is quite fast and a few units of time are sufficient to
yield
a collective change of opinion.

\begin{figure}
\centerline{\psfig{figure=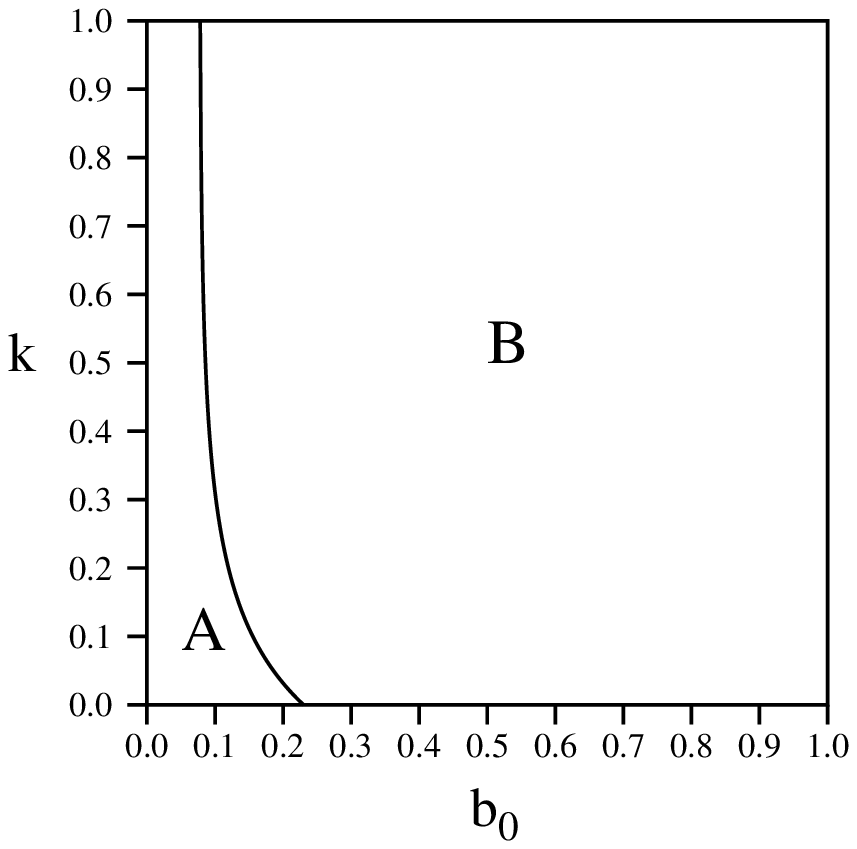,width=.45\textwidth}\hfill
             \psfig{figure=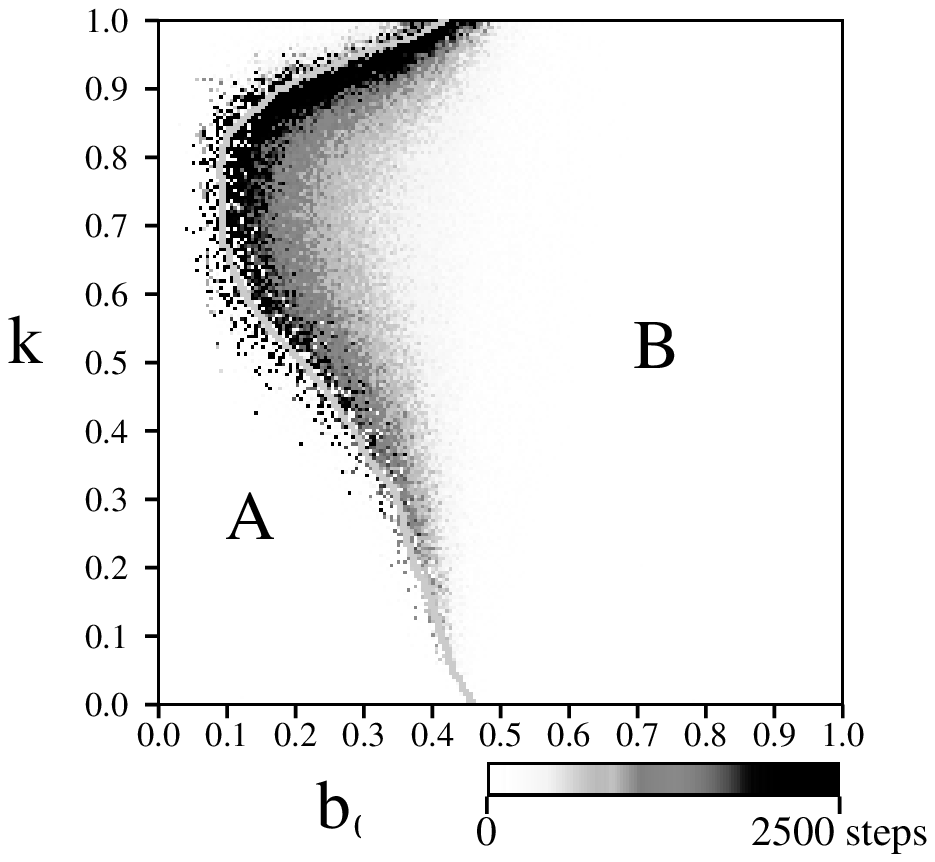,width=.45\textwidth}   }
\caption{left: Phase diagram for the model, with
$\beta=1$.
The curve delineates the regions where, on the left, $A$ wins with high
probability and, on the right, $B$ wins with probability one. The
outcome
depends on $b_0$, the initial density of $B$ and $k$, the probability of
a
confrontation.
Right: Same as the left panel but for a bias computed according
to the $B$ density on a local neighborhood of size $\ell=7$. The gray
levels indicate the time to eliminate the defeated species 
(dark for long time).}
\end{figure}

Following the same methodology, more complicated interactions between
individuals can be investigated. The case of a non-constant bias is
quite interesting and is described in~\cite{galam:98} and illustrated in Fig. 3
(left). In this latter case, the bias decreases locally if in a
neighborhood of diameter $\ell$ there are enough $B$.  There is a
re-entrance phenomena which shows that aggressivity, if too large, can
then be detrimental to the $B$ species.

\section{Finite size effects}

In this section we  demonstrate the essential role played by
finite size systems in the context of the present model and we
show that our model can be described in terms of a probabilistic phase
diagram which reduces to a trivial situation when the system size goes
to infinity. A more detailled analysis in given in~\cite{chopard:00}.

A possible conclusion is that some
socio-economical systems may be characterized by a strong sensitivity
to system size. For instance, the macroscopic behavior may change
dramatically whether the system is just large or if it is almost infinite.

The reason of this peculiar property is the existence, in such
systems, of statistically very rare configurations which drive the
evolution in a new and atypical way.  The observation that rare events
can develop and reach a macroscopic size has already been noticed in
other contexts. Examples are given by the generalized prisoner dilemma
problems~\cite{axelrod,antal,delahaye} or the recent work by
Solomon~\cite{solomon:00}. Percolation problems ~\cite{7} give another example
where a qualitative change of behavior is observed in the limit of an
infinite system~\cite{Adler:91}.

To illustrate this behavior, we consider a 1D system in which the effect
is more pronounced. The rule of the dynamics is a straightforward
variation of the above 2D case. We still consider four individuals per
cell and in order to conform to the topology restriction we change the
motion rule as follows: two individual randomly chosen among the
four travel to the left while the two others travel to the right.

\begin{figure}
\centerline{\psfig{figure=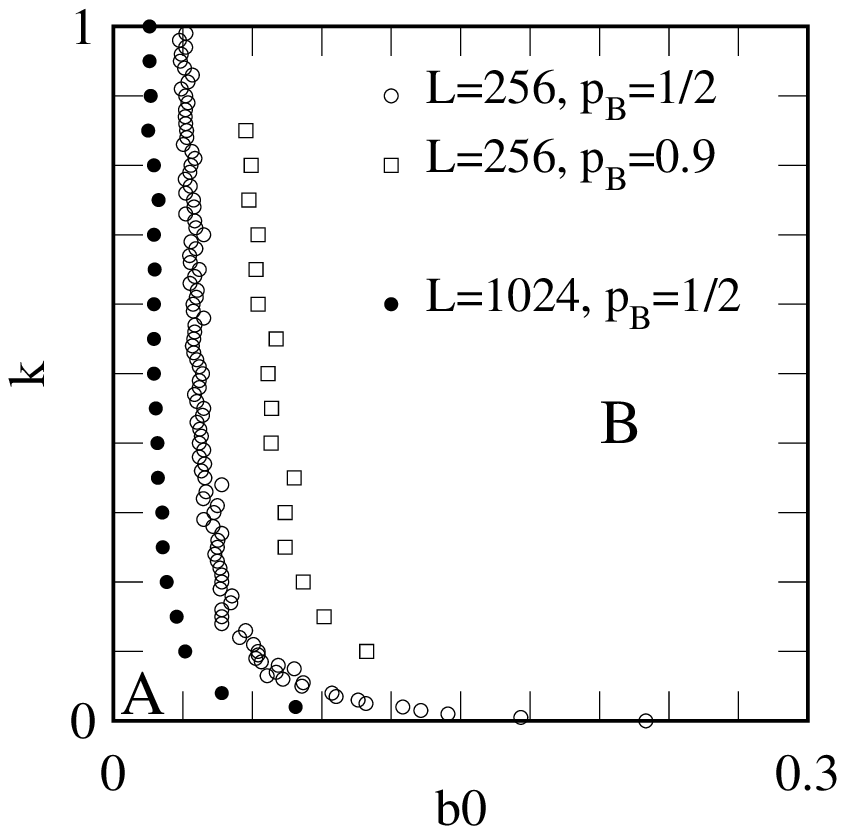,width=.4\textwidth}\hfill
            \psfig{figure=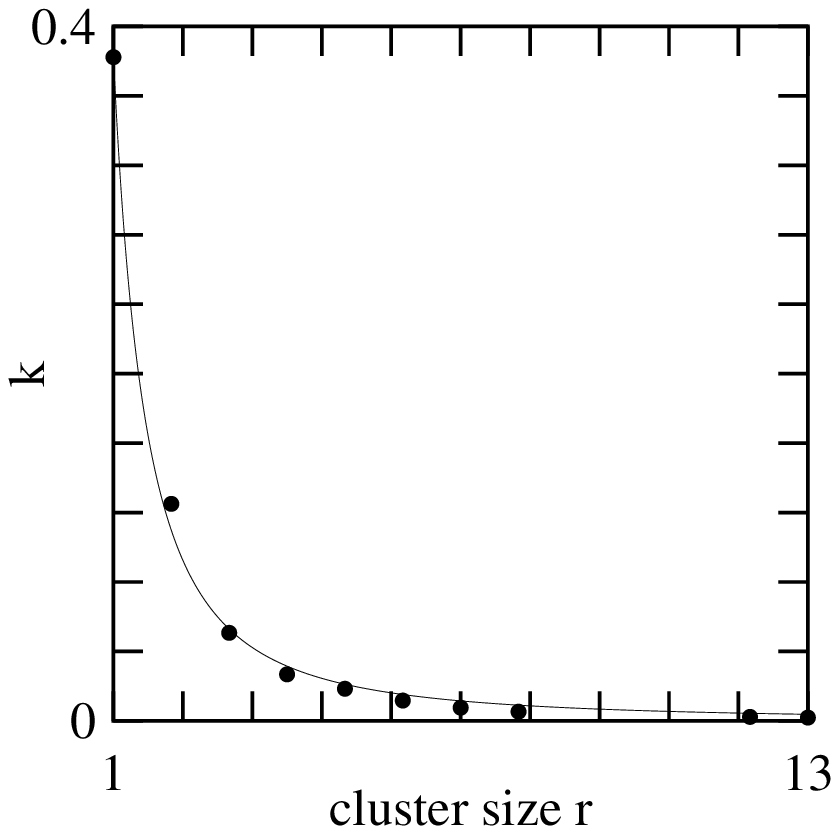,width=.4\textwidth} }
\caption{Left: Probabilistic stationary state phase diagram for a
systems of size $L=256$ and $L=1024$. Contour lines for $p_B=0.5$
and/or $0.9$ are shown. The region marked $B$ indicates that $p_B$ is
large whereas it is small in region $A$.  
Right: Critical size $r$ of
a single $B$ cluster that invades the system with probability $0.9$,
as a function of the aggressiveness $k$. Dots are the results of the
CA model and the solid line is an empirical fit: $k=1/(r^{1.8})$.}
\end{figure}

Here we study systems of linear size $L$ with periodic boundary
conditions.  For given values of $b_0$ and $k$ the dynamics is
iterated until a stationary state (either all $A$ or all $B$) is
reached. The interesting point is that the outcome of this experiment
is found to be probabilistic: the final state is all $B$ with
probability $p_B$ and all $A$ with probability $1-p_B$. Also, the
value of $p_B$ depends crucially on the system size $L$. As we shall
see, when $L\to\infty$, $p_B$ is one for the all $(b_0,k)$ plane.

For this reason, a standard phase diagram cannot describe the
situation properly. Thus, we propose a description in terms of what
we call a probabilistic phase diagram: each point of the $(b_0,k)$
plane is assigned a probability $p_B$ that the final state is entirely
$B$. Ideally, this diagram should be represented as a 3D plot. Instead,
in Fig.~4 (left), we show contour lines corresponding
to  given
probabilities $p_B$. Note that for the same value of $p_B$, the isoline
is shifted to the left as the system size increases.

These data show that if the aggressiveness $k$ is large enough,
initial configurations with a quite low density of $B$'s are able to
overcome the large initial majority of the $A$ species. The reason being
the presence of $B$ actors which are organized in small clusters such
that the diffusion is not effective enough to destroy them.  They
expand at a rate which makes them win systematically in the fights
against $A$ actors. Fig. 4 (right) is obtained by
considering a unique initial $B$ cluster of size $r$ in a sea of $A$'s. The
plot shows, for each value of $k$ the critical value of $r$ which
ensures that the $B$ cluster will invade all the system with
probability 0.9.

The result of Fig.~4 (right) is independent of the
system
size $L$ and the question is then how often such clusters
appear by chance. In a finite size system, with a given random
concentration $b_0$ of $B$ actors, there is always a finite
probability for such small clusters to exist in the initial
configuration. When this is the case the system will reach a pure $B$
stationary state. The larger the $L$ the more likely it is to observe
such a devastating cluster.

The way the separation line in Fig.~4 (left) depends on $L$ has been
investigated in Fig.~5 (left). The plot shows the location of the
transition line as a function of $L$ for a fixed probability $p_B=1/2$
and different values of $k$.

\begin{figure}
\centerline{\psfig{figure=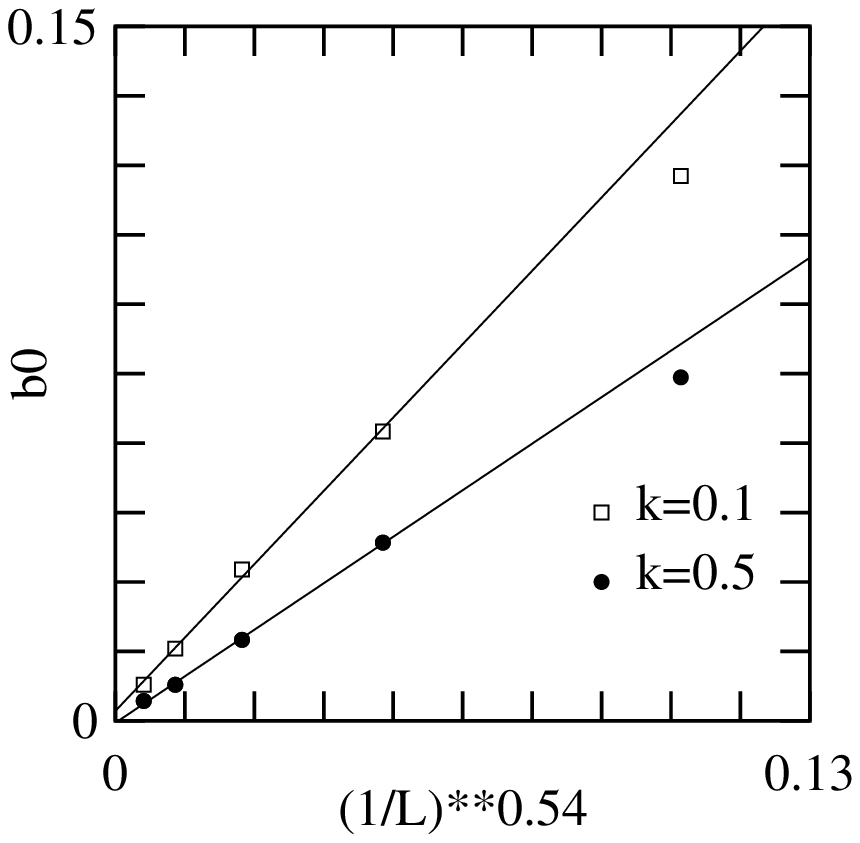,width=.4\textwidth}\hfill
            \psfig{figure=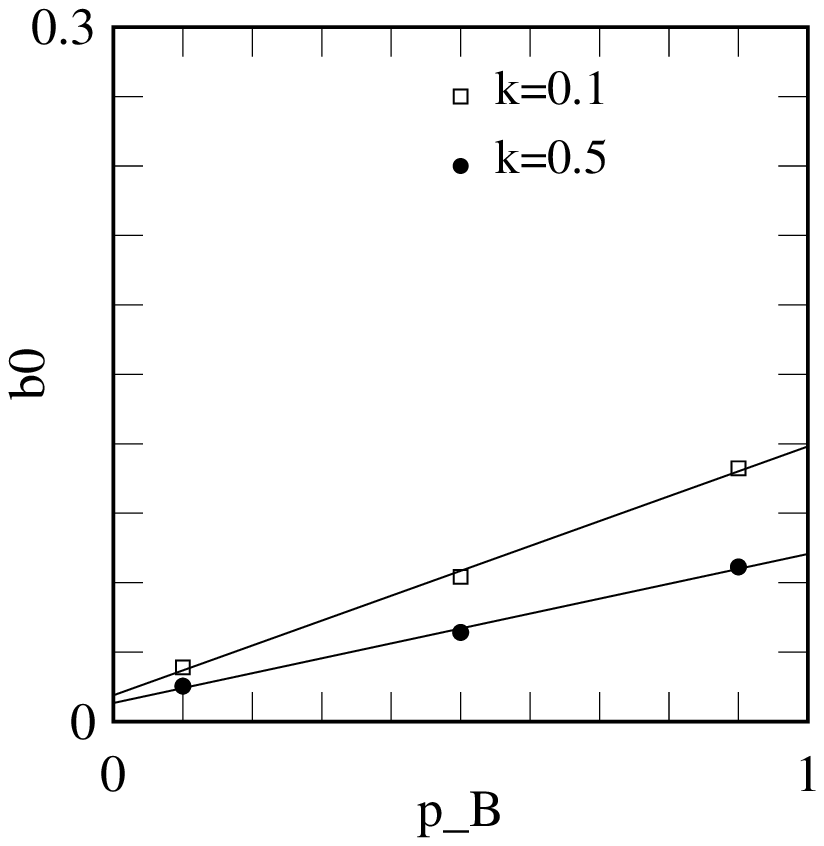,width=.4\textwidth}}
\caption{Left: Dependence of the critical density $b_0$ of $B$
particle as a function of the system size $L$, for a wining
probability $p_B=0.5$ and two values of $k$. We see that the $A-B$
separation line moves as $1/(L^{0.54})$.  Right: Critical initial
density $b_0$ as a function of $B$'s probability to win, $p_B$, for
two values of $k$ and $L=256$. From the assumption of a linear
dependence, the value of $b_0$ for $p_B=1$ can be interpolated. }
\end{figure}

One sees that when $L$ increases, the probabilistic line corresponding
to a given probability $p_B$ moves to the left and an extrapolation to
an infinite size system leads to a collapse of the transition line
with the vertical axis for all values $k \not =0$. For $k=0$, 
one recovers the mean-field transition point $b_{0c}=0.23$, for all
values of $p_B>0$. This is shown in
Fig.~4 for the case $L=256$, $p_B=1/2$, and can be
confirmed by
direct numerical simulations at $k=0$ (complete mixing of the individual
at each
time step).

These results show that the respective behaviors of finite size and
infinite size systems are qualitatively different.
Fig. 5 (right) shows, for a fixed system size $L=256$,
how the
critical density $b_0$ varies with $p_B$.  For two values of $k$, the
plot suggests an almost linear dependence.

In~\cite{chopard:00} we discuss in more detail the question of the
appearance of the
devastating $B$ clusters, that is 
the probability $P_L^{(r)}$ to find at least one cluster formed of
$r$ consecutive $B$ particles in a system of size $L$ providing that
the sites are randomly filled respectively with $B$ particles with
probability $b_0$ and with $A$ particles with probability $a_0=1-b_0$.
The result is that $P_L^{(r)}\to 1$ as $L\to\infty$, as long as $a_0\ne
0$.

\section{Conclusion}

In conclusion, although this model is very simple, it abstracts the
complicated behavior of real life agents by capturing some essential
ingredients. For this reason, the results we have presented may shed
light on the generic mechanisms observed in a social system of opinion
making.
In particular we see that the correlations existing between
successive fights may strongly affect the global behavior of the
system and that an organization is the key feature to obtain a definite
advantage over the other population. This
observation is important. For instance, during a
campaign against smoking or an attempt to impose a new system, it is
much
more efficient (and cheaper) to target the effort on  small nuclei
of persons rather than sending the information in an uncorrelated
manner.

\end{document}